\documentclass{jltp}
\usepackage{psfig}

\title{Thermal  conductivity via magnetic excitations in spin-chain materials}

\author{A. V. Sologubenko, T. Lorenz, H. R. Ott$^*$, and  A. Freimuth}

\address{II. Physikalisches Institut, Universit\"{a}t zu
K\"{o}ln, 50937 K\"{o}ln, Germany\\
$^*$Laboratorium f\"ur Festk\"orperphysik, ETH H\"onggerberg,
CH-8093 Z\"urich, Switzerland}

\runninghead{A. V. Sologubenko \textit{et al.}}{Thermal conductivity
in spin-chain materials}

\begin{document}

\maketitle

\begin{abstract}

We discuss the recent progress and the current status of
experimental investigations of spin-mediated energy transport in
spin-chain and spin-ladder materials with antiferromagnetic
coupling. We briefly outline the central results of theoretical
studies on the subject but focus mainly on recent experimental
results that were obtained on materials which may be regarded as
adequate physical realizations of the idealized theoretical model
systems. Some open questions and unsettled issues are also
addressed.

\end{abstract}

\section{INTRODUCTION}

Heat transport in solids is often very sensitive to even weak
disorder and is significantly influenced by phase transitions.
That is why measurements of the thermal conductivity are an
efficient experimental tool  in solid state physics. Several
books and review articles were devoted to thermal conduction in
solids.\cite{BermanBook,ZimanBook,Klemens58,Slack79} These
previous  authors usually discussed the heat transport by two
species of itinerant quasiparticles only, i.e., by considering
the  thermal conductivity by phonons  ($\kappa_{\rm ph}$) and by
itinerant charge carriers ($\kappa_e$). Even when dealing with
magnetic systems, the influence of magnetic excitations is
typically seen in a reduction of the phonon and electron thermal
conductivities via scattering of these quasiparticles by magnons
or magnetic impurities. The heat transport via the spin system
itself was, to a large extent, neglected. One of the main reasons
for this imbalance is that usually, it is very difficult to
separate the spin contribution ($\kappa_s$) from the total
measured thermal conductivity which, in most cases, is dominated
by phonon heat transport. This problem is particularly severe for
antiferromagnetic (AFM) three-dimensional (3D) materials where the
linear dispersion of the spin excitations causes, at low enough
temperatures, $\kappa_s$  to adopt a $T^3$ dependence, the same
as  $\kappa_{\rm ph}(T)$. A much more favorable situation is met
in strongly anisotropic spin systems, such as spin-chain
materials, where the exchange interaction along a particular
direction is much stronger than perpendicular to it. Since no
thermal conductivity via spin excitations is expected
perpendicular to the chain direction, investigating the
anisotropy of the heat transport is an efficient method for the
identification of $\kappa_s$. What really makes spin-chain
materials very attractive objects for transport measurements is,
however, the influence of quantum effects on their properties,
which are most pronounced for low spin values ($S=1/2$ and $S=1$).

Unusual transport of energy and magnetization in certain 1D
quantum spin systems was predicted long
ago.\cite{Huber69,Huber69a,Niemeijer71,Krueger71} More recent
theoretical studies along these lines induced a revival of
interest in unusual transport properties of integrable spin
systems.\cite{Castella95,Saito96,Zotos97,Narozhny98} During the
last decade, many new results have been obtained for transport
properties of various 1D quantum spin models, not only for the
''simple'' case involving an isotropic nearest-neighbor
interaction, but also taking into account higher-order neighbor
interactions, frustration effects and coupling to phonons. In
addition, more complex 1D spin arrangements such as ladders and
zig-zag chains were considered. The resulting theoretical
progress was reviewed in recent articles
\onlinecite{Zotos04_BookCh,Zotos05_Rev}. This progress, linked to
recent advances in preparation of materials whose features seem
to fulfill some of the assumptions of the various idealized
models quite well, has led to a considerable activity in
experimental studies of heat conduction in 1D spin systems. A
corresponding summary of earlier activities was given in
Ref.~\onlinecite{Sologubenko04_BookCh}.

In this article, we present a brief survey of recent experimental
results on the thermal conductivity via excitations in 1D AFM
spin systems. Heat transport in the 3D ordered spin state caused
by weak yet unavoidable interchain interactions is not discussed
here since it allows for a conventional description in terms of
well-defined quasiparticles (magnons). We do not attempt to
provide a full account of all the published papers on thermal
conductivity of spin-chain materials, but rather concentrate on
several central topics where some experimental progress has
recently been achieved. First, we discuss whether the observed
thermal transport in spin-chain compounds indeed reflects the
expectations that emerged from the mentioned theoretical work.
Then, we discuss what type of information on perturbations
affecting the spin-mediated transport can be extracted from
recent experimental data. Since most of the model materials are
insulators, the dominating influence is due to the interaction of
the spin system with phonons as well as with magnetic defects.
Another issue that we address here is how the predicted relations
between thermal conductivity and spin conductivity in
low-dimensional spin systems were attempted to be verified in
recent experiments. Finally, we discuss some features of the
thermal conductivity of spin chains in external magnetic fields.

\section{THERMAL TRANSPORT IN SPIN CHAINS AND LADDERS}

\subsection{Is it anomalous?}

In the modern theoretical literature, the energy transport in 1D
spin systems has been addressed employing either the Boltzmann
transport equation formalism or the linear response formalism.
The former approach, which relies on the notion of quasiparticle
modes with associated velocities and relaxation times, provides
transparent results and is well suited for the analysis of
experimental data. However, this quasiparticle picture is not
always applicable for quantum many-body systems. Thus, the most
widely used approach was directed towards the calculation of
transport coefficients via time-dependent current-current
correlation functions, due to Kubo.\cite{Kubo66} For the thermal
conductivity at a finite frequency  $\omega$,\cite{MahanBook90}
\begin{equation}
\label{Kappa_vs_w}
\kappa(\omega)= \frac{1}{T}\int_{0}^{\infty}
{dt e^{-i \omega t}     \int_{0}^{1/T}{d \tau
\langle   j_{\rm th}(-t-i\tau) j_{\rm th}     \rangle   }     },
\end{equation}
where $ j_{\rm th}$ is the energy current and $\langle ... \rangle$ denotes the thermodynamic average.
The real part of Eq.~(\ref{Kappa_vs_w}) can be decomposed into
\begin{equation}
\label{ReKappa}
{\rm Re}\,\kappa(\omega) = D_{\rm th} \delta(\omega) + \kappa_{\rm reg}(\omega),
\end{equation}
where the weight of the singular part ($D_{\rm th}$) is the
so-called thermal Drude weight. The experimentally accessible
quantity is the $dc$ conductivity ${\rm Re }\,\kappa(\omega \to
0) $.  A nonzero $D_{\rm th}$ implies a non-decaying energy
current and, thus, ballistic heat transport. Since the energy
current is one of the  conserved
quantities,\cite{Castella95,Zotos97,Naef98} the existence of a
nonzero Drude weight is often linked with the integrability of a
system. The question whether $D_{\rm th}$  may, for various spin
systems, adopt a nonzero value and how it evolves under the
influence of different perturbations are the central points of
current theoretical studies of thermal transport in quantum spin
systems. The thermal Drude weight $D_{\rm th}(T)$ is as important
for thermal transport as is the Drude weight  $D(T)$ for electric
transport where its value at $T=0$ distinguishes an ideal
conductor ($D(0)>0$) from an insulator($D(0)=0$), as was put
forward by Kohn.\cite{Kohn64} Extending   Kohn\/'s conjecture to
heat transport and to nonzero temperatures,\cite{Castella95}
several cases can be distinguished:
\\
(a) thermal insulators if $D_{\rm th}(T)=0$ and $\kappa_{\rm reg}(\omega \to 0,T)=0$;\\
(b) conventional thermal conductors if $D_{\rm th}(T)=0$ and $\kappa_{\rm reg}(\omega \to 0,T) > 0$;\\
(c) ideal thermal conductors if $D_{\rm th}(T)>0$ for any value of $\kappa_{\rm reg}(\omega \to 0,T)$.\cite{Zotos04_BookCh}

However, an experimentally observed, anomalously large $dc$
thermal conductivity does not necessary imply a ballistic thermal
transport (case (c)). It is also possible that, even if  $D_{\rm
th}(T)=0$ (case (b)), the low-$\omega$ part of $\kappa_{\rm
reg}(\omega,T)$ may adopt anomalously large values for certain
systems with slowly decaying energy currents. Even if
calculations for an idealized pure system suggest the existence
of a nonzero Drude weight, in any real system, perturbations such
as phonons, 3D couplings, or defects may destroy the
integrability of the system. As a result, the $\delta$-peak
broadens into a Lorentzian with a width inversely proportional to
the relaxation time $\tau(T)$, the value of which depends on the
type and strength of the perturbation, and the thermal
conductivity is reduced accordingly. Thus, only a comparison of
experimental data with theoretical calculations, taking into
account both the singular and the regular parts of the
conductivity spectrum, provides a reliable judgement of the
situation.

Real materials of current interest which may be regarded as
reasonable physical realizations of model low-dimensional spin
$S$ systems belong to three classes: $S=1/2$ chains, $S=1$
chains, and two-leg $S=1/2$ ladders. An often encountered case is
the Heisenberg $S=1/2$ XXZ chain, for which the Hamiltonian is
\begin{equation}
\label{eHamiltonian12}
H=J \sum_{i} (S_i^x S_{i+1}^x + S_i^y S_{i+1}^y + \Delta S_i^z S_{i+1}^z) ,
\end{equation}
where $J>0$ is the intrachain nearest-neighbor exchange coupling
and $\Delta$ characterizes the anisotropy of the interaction. The
excitations are pairs of $S=1/2$ quasiparticles (spinons) and the
spectrum consists of a continuum with an upper and a lower
boundary. It is gapless for $| \Delta | \leq 1$ but a spin gap
opens for larger $| \Delta | $. The isotropic case  $\Delta=1$ is
especially important because a number of model systems with  $J$
ranging from quite high (e.g., Sr$_2$CuO$_3$ with $J/k_B \approx
2000$~K)\cite{Motoyama96} to rather low values (CuPzN with $J/k_B
\approx 10$~K)\cite{Hammar99} have  been realized in appropriate
material syntheses. Systems which are described by
Eq.~(\ref{eHamiltonian12}) are integrable for $| \Delta | < 1$
(easy plane) as well as for $\Delta=1$ (isotropic). In several
theoretical papers and using different methodical approaches it
has unambiguously been demonstrated that at $T>0$, the thermal
Drude weight is nonzero if $| \Delta | \leq
1$.\cite{Kluemper02,HeidrichMeisner02,Orignac03,Sakai05_MTE}
Other work established the thermal transport in $S=1/2$ chains in
the gapped phases, such as XXZ chains with Ising-type anisotropy
$| \Delta | > 1$,\cite{Sakai03_XXZ} with next-nearest neighbor
interactions,\cite{Jung06} with three-spin
interactions,\cite{Lou04} with dimerization and
frustrations.\cite{HeidrichMeisner02,Orignac03,Saito02,Saito03}
For $T > 0$, some calculations predicted a nonzero $D_{\rm th}$
in the massive regime;\cite{Orignac03,Sakai03_XXZ,Saito03} other
studies question this conclusion for dimerized and frustrated
chains, however.\cite{HeidrichMeisner02,HeidrichMeisner03}

The same uncertainty about  $D_{\rm th}(T>0)$ exists for $S=1$
chain systems. The excitation spectrum of the isotropic AFM $S=1$
model has a large gap; the system is nonintegrable. Although
earlier calculations provided support for a nonzero thermal Drude
weight at all temperatures,\cite{Orignac03} more recent work
suggests that  $D_{\rm th}(T)$ vanishes in the thermodynamic
limit.\cite{Karadamoglou04} In the latter case, the thermal
conductivity of the spin system is expected to be completely
governed by the regular contribution $\kappa_{\rm reg}(\omega,T)$.

A great deal of recent theoretical interest concentrated on the
transport properties of two-leg $S=1/2$ ladders. This interest
was stimulated by experimental observations of a rather large
spin thermal conductivity along the ladder direction in variants
of (Ca,Sr)$_{14}$Cu$_{24}$O$_{41}$ which contains spin ladders as
a structural element.\cite{Sologubenko00_lad} The ladders are
$S=1/2$ chains (legs) connected in pairs, with an exchange
constant $J$ along the legs and $J_{\perp}$ perpendicular to the
legs (along the rungs). The interaction along the rungs is the
integrability-breaking perturbation because for $J_{\perp}=0$,
the system is just a set of $S=1/2$ chains as described above.
For $J_{\perp}>0$, the situation with the thermal Drude weight is
controversial, similar to other nonintegrable systems with spin
gaps mentioned above: the nonzero thermal Drude weight calculated
for this system disagrees with more recent numerical
calculations.\cite{Orignac03,HeidrichMeisner03,Alvarez02_Ano,Zotos04}

Measurements of the thermal conductivity were made for several
physical realizations of 1D spin systems. Before discussing the
experimental results for $\kappa_s$, we wish to mention some
aspects of ambiguity in the data analysis. The standard method of
measuring the components of the thermal conductivity tensor ${\bf
\kappa}$ relies on the Fourier law of proportionality
\begin{equation}\label{eFourier}
    {\bf J}_{\rm th}  = - {\bf \kappa} \nabla T,
\end{equation}
linking the heat flux ${\bf J}_{\rm th}$  and the temperature
gradient $\nabla T$. The heat flux is typically directed along
one of the main crystallographic axes, e.g. $\alpha$, therefore
we consider $\kappa_{\alpha\alpha} \equiv \kappa^{\alpha}$. The
measured thermal conductivity always includes contributions from
all delocalized excitations acting as heat carriers. Therefore,
separating the spin thermal conductivity $\kappa_s$ from
contributions of phonons $\kappa_{\rm ph}$, electronic
quasiparticles $\kappa_e$ etc., is, in general, not
straightforward. Although  $\kappa_e$ is negligibly small or zero
for most studied spin-chain compounds, the phonon contribution is
always larger or, at least, of the same order of magnitude as
$\kappa_s$. In principle, one may try to reduce $\kappa_{\rm ph}$
by introducing some lattice defects which scatter phonons much
stronger than spin excitations. This step has successfully been
applied to the spin-ladder compounds
(Sr,Ca,La)$_{14}$Cu$_{24}$O$_{41}$.\cite{Hess01} Unfortunately,
the deliberately introduced disorder produces some unavoidable
additional scattering of spin excitations as well.

The most widely used approach for establishing $\kappa_s$ exploits
the fact that, because of very weak interchain interactions, the
spin contribution to the transport perpendicular to the chains is
usually negligible. Thus, for establishing the spin contribution
one has to compare the thermal conductivity $\kappa^{\parallel}$
measured parallel to the chains (which contains both $\kappa_{\rm
ph}$ and $\kappa_s$) with the thermal conductivity
$\kappa^{\perp}$ perpendicular to the chains (which contains only
$\kappa_{\rm ph}$). Still, because of the possible anisotropy of
phonon transport, this method is, in some cases, only partly
successful in avoiding substantial uncertainties in the
evaluation of $\kappa_s(T)$. This is especially the case in the
vicinity of phase transitions, where the phonon scattering by
critical fluctuations is often very strong and not necessarily
isotropic. For example, the existence of thermal conduction via
the spin system was discussed for the $S=1/2$ chain cuprate
CuGeO$_3$.\cite{Ando98,Salce98,Takeya00,Takeya00b} Here the Cu
atoms are linked via superexchange with $J_c$=10.4~meV  along the
chain direction ($c$ axis), while in the perpendicular
directions, the coupling is at least an order of magnitude weaker
($J_{b} \sim 0.1 J$ and $J_{a} \sim -0.01 J$).\cite{Nishi94} The
temperature dependence of the thermal conductivity
$\kappa^{\parallel}(T)$ along the chain direction exhibits a
two-peak structure. It has been speculated that the
higher-temperature peak  is caused by the spin contribution to
$\kappa^{\parallel}(T)$, while the lower-temperature peak is of
phononic origin.\cite{Ando98,Salce98} However, the same features
are also observed for $\kappa^{\perp}(T)$, where $\kappa_s$ is
expected to be negligible.\cite{Salce98,Hofmann02} Taking into
account that the spin-Peierls transition occurs exactly between
the two peaks at $T_{SP} \approx 14 {\rm ~K}$, the two-peak
feature may safely be attributed to phonon transport with
enhanced scattering close to the
transition.\cite{Vasilev97,Vasilev98,Hofmann02} As a result,
despite the large amount of experimental work on thermal
transport in CuGeO$_3$, the identification of spin-carried heat
conduction in this material has not been established
unequivocally. A similar situation is met for the S = 1/2 chain
compounds KCuF$_3$ and BaCu$_2$Si$_2$O$_7$. Because of moderately
weak interchain interactions, the 3D magnetic ordering sets in at
relatively high temperatures (39.8~K and 9.2~K for KCuF$_3$ and
BaCu$_2$Si$_2$O$_7$, respectively) and is reflected in the
temperature dependences of $\kappa_{\rm ph}$ and
$\kappa_s$.\cite{Miike75,Sologubenko03_Uni}

\begin{figure}
\centerline{\psfig{file=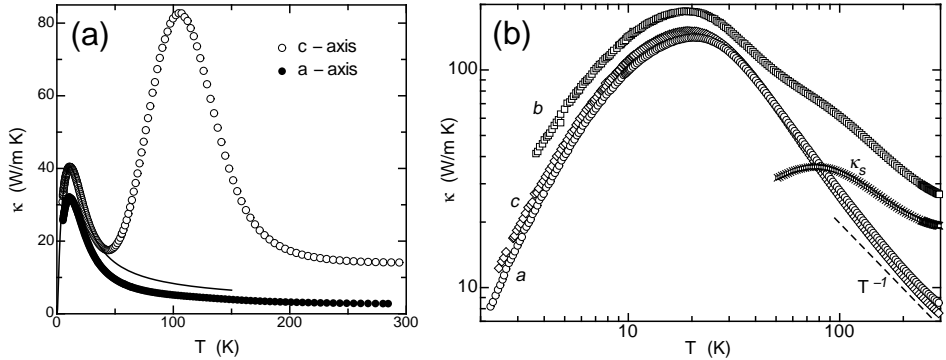,height=1.9in}}
\caption{ (a) Thermal conductivity of the spin-ladder compound
Sr$_{14}$Cu$_{24}$O$_{41}$.\cite{Sologubenko00_lad,Sologubenko00lp}
The ladders are parallel to the $c$ axis. (b) Thermal
conductivity of the $S=1/2$ chain material
Sr$_2$CuO$_3$.\cite{Sologubenko00_213} The chains run parallel to
the $b$ axis. } \label{LadChains}
\end{figure}

Only when the temperature dependences of the thermal
conductivities perpendicular to the chains are smooth and
featureless, they can be used for evaluating the phonon
background parallel to the chains. Examples are shown in
Fig.~\ref{LadChains}.  For the two materials, $\kappa^{\perp}(T)$
exhibits a single peak at low temperatures and, at elevated
temperature, $\kappa^{\perp}(T)$ varies approximately as $T^{-1}$
which is typical for phonon thermal conduction of non-magnetic
insulators.\cite{BermanBook} In turn, the thermal conductivity
parallel to chains $\kappa^{\parallel}(T)$ shows, on top of the
phonon contribution, a distinct additional feature at higher
temperatures. These extra contributions were attributed to
thermal transport via the spin system, represented by
$\kappa_s(T)\equiv\kappa^{\parallel}_s(T)$. The phonon background
$\kappa_{\rm ph}^{\parallel}(T)$ can be analyzed and subtracted
from $\kappa^{\parallel}(T)$ but it is obvious that the
uncertainties in $\kappa_s(T)$ are large for temperatures where
the phonon background dominates.

If we now turn to the experimental results so far obtained for
$\kappa_s(T)$, the main question is whether these results are
consistent with the ballistic heat transport expected for
integrable systems. If we define ballistic transport as transport
where the mean free path of the excitations is limited by the
sample dimensions, then the answer would be negative. The mean
free path $l_s$  of 1D spin excitations can roughly be evaluated
using the kinetic equation
\begin{equation}
\kappa_s(T)  = C_s(T) v_s(T) l_s(T),
\label{eCvl}
\end{equation}
where $C_s$ is the magnetic specific heat and $v_s$ is the mean
velocity of spin excitations. The data for $l_s(T)$ are available
for $S=1/2$ chain materials, such as
Sr$_{2}$CuO$_{3}$,\cite{Sologubenko00_213}
SrCuO$_{2}$,\cite{Sologubenko01}
BaCu$_2$Si$_2$O$_7$,\cite{Sologubenko03_Uni}
CuGeO$_{3}$,\cite{Ando98,Salce98,Takeya00}
Yb$_4$As$_3$,\cite{Schmidt01} as well as for the $S=1$ chain
materials AgVP$_2$S$_6$,\cite{Sologubenko03_AVPS} and
Y$_2$BaNiO$_5$,\cite{Kordonis06} and some variants of the $S=1/2$
two-leg ladder compound
(Ca,Sr,La)$_{14}$Cu$_{24}$O$_{41}$.\cite{Sologubenko00_lad,Hess01}
In all cases, $l_s$ was never larger than a few $10^3$~\AA,\/
much shorter than the lateral extensions of typical samples which
were of the order of 1~mm.

\begin{figure}
\centerline{\psfig{file=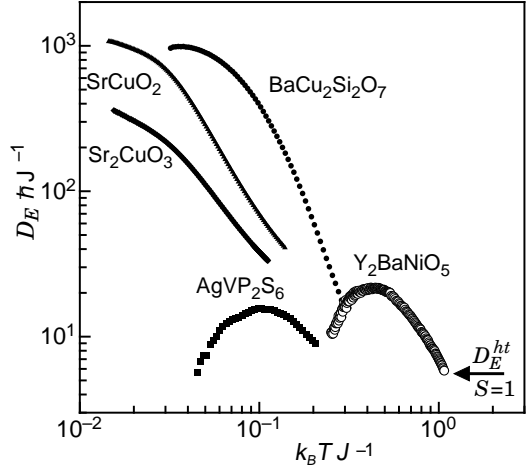,height=2.5in}}
\caption{The energy diffusion constant $D_{E}(T)$ calculated from
the thermal conductivity data of the $S=1$ chain compounds
AgVP$_2$S$_6$ and Y$_2$BaNiO$_5$, and of the  $S=1/2$ chain
compounds BaCu$_2$Si$_2$O$_7$  and SrCuO$_2$ (data from
Refs.~\onlinecite{Sologubenko03_Uni,Sologubenko01,Sologubenko03_AVPS,Kordonis06}).
The arrow corresponds to the high-temperature limit $D^{ht}_E$
calculated for $S=1$ chains in Ref.~\onlinecite{Karadamoglou04}.}
\label{DE}
\end{figure}

However, a more general definition of ballistic transport is
provided by the presence of non-decaying contributions to the
energy current which can be shown to be equivalent to the
existence of a nonzero Drude weight. This definition is more
appropriate because it does not rely on the validity of the
quasiparticle approach. Applying this criterion in analyzing the
experimental results of steady-state heat transport experiments,
the identification of the ballistic transport relies on
demonstrating that the measured $\kappa_s$ exceeds the values of
the thermal conductivity provided by the regular contribution in
Eq.~(\ref{ReKappa}). Experimental data available in the
literature indeed seem to support the theoretical predictions
which claim ballistic thermal transport in $S=1/2$ uniform
Heisenberg chains and diffusive transport in $S=1$ chains. In
Fig.~\ref{DE}, we plot the spin-related energy diffusion
constants $D_{E}$ that were calculated from experimental data of
$\kappa_s(T)$ for several spin-chain compounds\cite{Kordonis06}
by using $D_{E}(T) = \kappa_{s}(T) / (C_{s}(T) a^{2})$, where
$C_{s}(T)$ is the specific heat of the spin chain and $a$ the
lattice constant along the chains. For $S=1$ chains, the energy
diffusion constant has recently been calculated from the regular
contribution $\kappa_{\rm reg}(\omega,T)$ for high
temperatures.\cite{Karadamoglou04} The high-temperature limiting
value $D^{ht}_E$ is indicated by an arrow in  Fig.~\ref{DE}. For
both $S=1$ compounds presented in  Fig.~\ref{DE} the energy
diffusion constants are clearly of the same order of magnitude as
$D^{ht}_E$ values. In contrast, $D_E(T)$ for the $S=1/2$ chains
are much larger than for the $S=1$ species, they exhibit a strong
temperature dependence and do not scale with $T/J$, as one would
expect for diffusion that is governed by intrinsic interactions
only. Taking into account that the theory predicts $\kappa_{\rm
reg}(\omega,T)=0$ for $S=1/2$ chains in the gapless regime, this
comparison strongly suggests that intrinsic diffusion is the
dominating process in $S=1$ chains but much less so in $S=1/2$
chains. A comparison between calculated values of $\kappa_{\rm
reg}(\omega,T)$ and experimental results for $\kappa_s(T)$ has
recently been made for two-leg spin ladders.\cite{Zotos04} This
comparison suggests that the experimentally observed thermal
conductivity via the spin system is governed by intrinsic
diffusion.  The emerging understanding of the dichotomy between
diffusive and ballistic transport in 1D spin systems is that the
thermal transport in systems with a spin gap, such as $S=1$ chains
or $S=1/2$ even-leg ladders, is determined by intrinsic
diffusion, while for $S=1/2$ spin chains with gapless spectra,
the thermal transport is ballistic and limited by external
perturbations.

\subsection{Scattering mechanisms}

Although most of the relevant theoretical work was based on the
Kubo formalism, the analysis of experimental data on the thermal
conductivity of low dimensional spin systems is usually employing
Boltzmann\/'s kinetic transport theory. In this approximation

\begin{equation}
\kappa_s = \sum_{k} \frac{df(k,T)}{dT}\varepsilon(k) v_s(k) l_s(k,T),
\label{eKinEq}
\end{equation}
where $f$, $\varepsilon$, $v_s$, and $l_s$ are, respectively, the
distribution function, the energy, the velocity, and the mean
free path of spin excitations with wavevector $k$. The mean free
path $l_s$ is related to the relaxation time $\tau$ as $l_s = v_s
\tau$.  Assuming that several scattering mechanisms act
independently of each other, in most cases a good approximation
for the total mean free path is
\begin{equation}
l_{s}^{-1}(k,T)=\sum_i l_{s,i}^{-1}(k,T),
\label{eLs}
\end{equation}
where each $l_{s,i}(k,T)$ term corresponds to an independent
scattering channel. In order to keep the analysis tractable, in
most experimental papers it is usually assumed  that the
scattering rates are  $k$-independent, such that $
l_{s,i}(k,T)\equiv l_{s,i}(T)$. This assumption is often
questionable but largely dictated by the lack of properly
developed theoretical formalisms describing the relevant
scattering processes of spin excitations in 1D systems.
Nevertheless, it is very useful in the sense that it allows to
use the simplified expression of  Eq.~(\ref{eCvl}) to estimate
the values and the temperature dependence of the average mean
free path of excitations in the corresponding spin system.

\begin{figure}
\centerline{\psfig{file=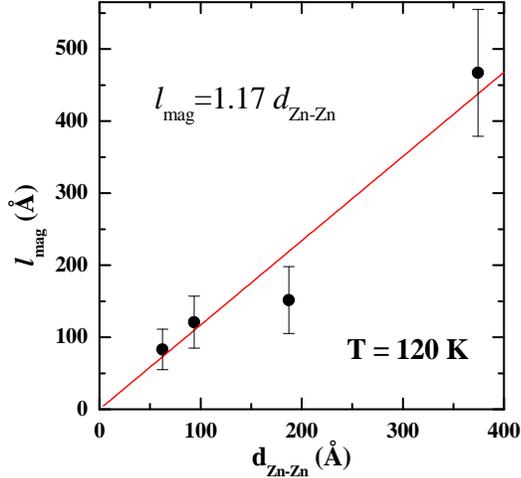,height=2.5in}}
\caption{ Magnon mean free path as a function of the mean
distance between Zn ions along the ladders $d_{\rm Zn-Zn}$ in
Sr$_{14}$Cu$_{24-x}$Zn$_x$O$_{41}$ (from
Ref.~\onlinecite{Hess06}).} \label{Hess06fig3}
\end{figure}

In many experimental publications on the spin thermal conductivity
in 1D spin chains and
ladders,\cite{Takeya00,Sologubenko00_lad,Sologubenko01,Hess01,Schmidt01,Kudo01_ZnNi,Sologubenko03_Uni,Kordonis06,Takeya00b,Hess04_lad,Hess05,Ribeiro05,Hess06}
magnetic defects were considered as major scattering centers for
spin excitations. It was always assumed that the corresponding
scattering rates and hence the mean free paths ($l_{s,{\rm
def}}$) were  $T$- and $k$-independent. In several studies, the
influence of intentionally introduced disorder on the mean free
path of spin excitations was
investigated.\cite{Takeya00,Takeya00b,Kudo01_ZnNi,Ribeiro05,Hess06}
It was shown that the reduction of  $l_{s,{\rm def}}$ indeed
correlates with the concentration of magnetic defects, as is
illustrated in Fig.~\ref{Hess06fig3} for the spin-ladder compound
Sr$_{14}$Cu$_{24}$O$_{41}$ with Zn serving to partially replace
Cu on the corresponding sublattice.\cite{Hess06} Analogous
correlations of the values of $l_{s,{\rm def}}$ with the
concentration of defects, estimated from NMR and magnetic
susceptibility measurements, were reported in
Refs.~\onlinecite{Sologubenko00_213,Kordonis06}.

There is no doubt that magnetic defects are very effective
scatterers of spin excitations in 1D magnetic systems. However,
the convenient assumption that $l_{s,{\rm def}}$ is $T$- and
$k$-independent, seems not to be justified in general. The
scattering of spin excitations by impurities and disorder was
addressed theoretically in
Refs.~\onlinecite{Rozhkov05,Chernyshev05} for gapless $S=1/2$
chains. The calculations suggest that the impurity relaxation
time (and thus $l_{s,{\rm def}}$ ) should be proportional to
temperature. In Ref.~\onlinecite{Orignac03}, the scattering by
impurities was analyzed for 1D spin systems with a gap in the
excitation spectrum. It was found that the lifetime of spin
excitations is energy dependent for the single-impurity case, and
the situation is even more complicated for higher concentrations
of impurities.

Another species of scatterers needs to be taken into account in
analyzing the data for the spin-ladder compound
(Ca,Sr)$_{14}$Cu$_{24}$O$_{41}$ where, depending on the chemical
composition, different concentrations of mobile holes are present
in the ladders. In Ref.~\onlinecite{Sologubenko00_lad} it was
suggested that in Sr$_{14-x}$Ca$_{x}$Cu$_{24}$O$_{41}$  ($x=0,
2$) the holes are the principal scatterers of magnons at high
temperatures. This conclusion was based on the observation that
$l_s$ correlates with the average distance between holes.
Subsequent experiments on (Ca,Sr)$_{14}$Cu$_{24}$O$_{41}$
provided additional supporting evidence for the magnon-hole
scattering in this material.\cite{Hess01,Hess04_lad} The holes
are considered as a kind of mobile defects which scatter magnons
in the same way as static defects, thus providing a $T$- and
$k$-independent term to the total inverse mean free path if the
concentration of holes does not vary with temperature. At low
temperatures, this assumption is not valid because the holes
either adopt an ordered configuration or are transferred out of
ladders.\cite{Hess04_lad}

At not very low temperatures, the scattering of spin excitations
by phonons is expected to be significant. This scattering
mechanism was taken into account in the analyses of experimental
results for several spin-chain and spin-ladder materials by
including a term $l_{s,{\rm ph}}^{-1}(T)$ on the right-hand side
of
Eq.~(\ref{eLs}).\cite{Hess01,Sologubenko03_Uni,Sologubenko00_213,Sologubenko01,Hess05}
Phenomenological expressions for $l_{s,{\rm ph}}^{-1}(T)$ were
suggested for spinon-phonon Umklapp processes in $S=1/2$
chains\cite{Sologubenko03_Uni,Sologubenko01} and two magnon - one
phonon scattering events in two-leg $S=1/2$ ladders.\cite{Hess01}
A theory for the spin thermal conductivity in $S=1/2$ chains,
taking into account spinon-phonon Umklapp scattering, was
developed in Ref.~\onlinecite{Shimshoni03}. In this work, the
transport in integrable systems with perturbations that weakly
violate the conservation laws is considered. The presence of both
spinon-spinon Umklapp processes and spinon-phonon scattering was
demonstrated to lead to an exponential decrease of $\kappa_s(T)$
with increasing temperature, in apparent agreement with
$\kappa_s(T)$ calculated from experimental data for Sr$_2$CuO$_3$.
Spinon-phonon scattering in combination with impurity scattering
in  $S=1/2$ chains was theoretically analyzed in
Refs.~\onlinecite {Rozhkov05,Chernyshev05}. Three regimes with
different temperature dependences for the thermal conductivity
were predicted: At low temperatures it is expected that
$\kappa_s(T) \propto T^2$. At intermediate temperatures,
$\kappa_s(T) \propto 1/T$, and at high temperatures,  $\kappa_s$
is expected to be constant. The experimental values for
$\kappa_s(T)$ of Sr$_2$CuO$_3$, presented in
Ref.~\onlinecite{Sologubenko00_213}, seem to approach the
predicted high-temperature constant value but, unfortunately, the
temperature range covered by these measurements does not extend
to high enough values in order to verify the predicted $\kappa_s=
const$ satisfactorily.

The role of phonons is not restricted to serving as a reservoir
for the exchange of energy and momentum with the quasiparticles
of the spin system in the related scattering processes. First of
all, a non-vanishing interaction between the lattice and the spin
system is necessary for the very observability of $\kappa_s$ in a
typical thermal conductivity experiment, where the heat flux is
first introduced into the phonon subsystem and then redistributed
between phonons and spin excitations.\cite{Sanders77} Since in an
experiment probing the thermal conductivity, the spin excitations
interact with phonons which are not in equilibrium but have a
nonzero net momentum along the temperature gradient, these
interactions lead to a redistribution of momentum between phonons
and spin excitations. If the scattering between phonons and spin
excitations dominates over all other processes which lead to
scattering within the spin systems (defects and
Umklapp-processes), the resulting drag effects may modify
$\kappa_s$. The influence of such drag effects on the thermal
conductivity of conventional (3D) FM and AFM systems was
discussed in Refs.~\onlinecite{Gurevich66,Gurevich67}. Depending
on the relative strengths of magnon-phonon, magnon-defect, and
phonon-defect scattering, and on the type of defects (magnetic or
nonmagnetic), complex $\kappa(T)$ variations were predicted.
Recently, the problem of spin drag has theoretically been
addressed for spin ladders.\cite{Boulat06cm} It was shown that
spin-phonon drag effects modify both $\kappa_{\rm ph}$ and
$\kappa_s$ and also cause interference effects which are taken
into account by the introduction of $\kappa_{s,{\rm ph}}$. With
respect to experiment, no comparison of experimental data with
spin-phonon drag predictions have been published up to now.

\subsection{ Influence of a magnetic field}

According to recent theoretical work, an external magnetic field
can modify $\kappa_s$ not only because it changes the spectrum of
magnetic excitations but also via so-called magnetothermal
corrections.\cite{Louis03,HeidrichMeisner05} In general, the heat
current $j_E$ and the magnetization current  $j_M$ are related to
the temperature gradient $\nabla T$ and the magnetic field
gradient $\nabla B$ by
\begin{equation}\label{eLij}
\left(\begin{array}{c}j_M\\ j_E\end{array}\right)=
\left(\begin{array}{cc} L_{MM}&L_{ME}\\
L_{EM}& L_{EE}\end{array}\right)
\left(\begin{array}{c}\nabla B\\
-\nabla T\end{array}\right).\end{equation} In zero magnetic
field, $L_{EM} = L_{ME}=0$ and hence, the coefficient $L_{EE}$ is
equivalent to the thermal conductivity $\kappa_s$ (in the
following only the magnetic contribution to the heat transport is
considered). In magnetic fields, the off-diagonal elements in
(\ref{eLij}) may be nonzero, and this leads to a number of
interesting effects. For example, by analogy with the Seebeck
effect in electric conductors, where the heat flux induces an
electric voltage along the direction of the flux, $j_E$ is
expected to induce a gradient of magnetic induction in magnetic
systems. A number of interesting predictions for magnetothermal
effects in $S=1/2$ chain compounds which fulfill  the XXZ-model
conditions, can be found in recent work in
Refs.~\onlinecite{Louis03,Sakai05_MTE,Furukawa05}. At present, no
experimental results on magnetothermal effects in spin chains are
available, mainly because of considerable practical difficulties
in carrying out such experiments.\cite{Louis03}

Nonzero off-diagonal elements in (\ref{eLij}) also modify the
thermal conductivity $\kappa_s$ of the spin
system.\cite{Sakai05_MTE,Sakai03_XXZ,HeidrichMeisner05} Since
heat transport experiments are normally done under the condition
of $j_M=0$, it follows from the definition of $\kappa$ in
Eq.~(\ref{eFourier}) that, for $B \neq 0$,
\begin{equation}\label{eKsLij}
\kappa_s = L_{EE} - \frac{1}{T}\frac{L_{EM}^2}{L_{MM}}.
\end{equation}
The second term on the right-hand side of Eq.~(\ref{eKsLij})
represents the magnetothermal correction, which was theoretically
investigated in detail in
Refs.~\onlinecite{Sakai05_MTE,HeidrichMeisner05}. Nontrivial
dependences of these corrections on temperature, magnetic field
and the anisotropy parameter $\Delta$ in (\ref{eHamiltonian12})
were predicted. Unfortunately, an experimental evidence for
magnetothermal corrections is still missing. Most of the good
physical realizations of the 1D spin model systems discussed
above, such as Sr$_2$CuO$_3$, are not suitable for this type of
experiments, because of the extremely large exchange constants
with $J/k_B$ of the order of 10$^2$-10$^3$~K. Available
laboratory magnetic fields are much too small to influence such
systems in any significant way. Moreover, Shimshoni {\em et
al.}\cite{Shimshoni03}  argued against the existence of these
corrections in real materials because the conservation of the
total magnetization may be broken by, e.g., spin-orbit coupling.
In their paper, also the influence of magnetic field on the spin
thermal conductivity of $S=1/2$ chains was addressed. Complex
fractal-like dependences of $\kappa_s$ on the magnetization,
resulting from field-induced variations of spin-phonon Umklapp
scattering, were predicted.

\begin{figure}
\centerline{\psfig{file=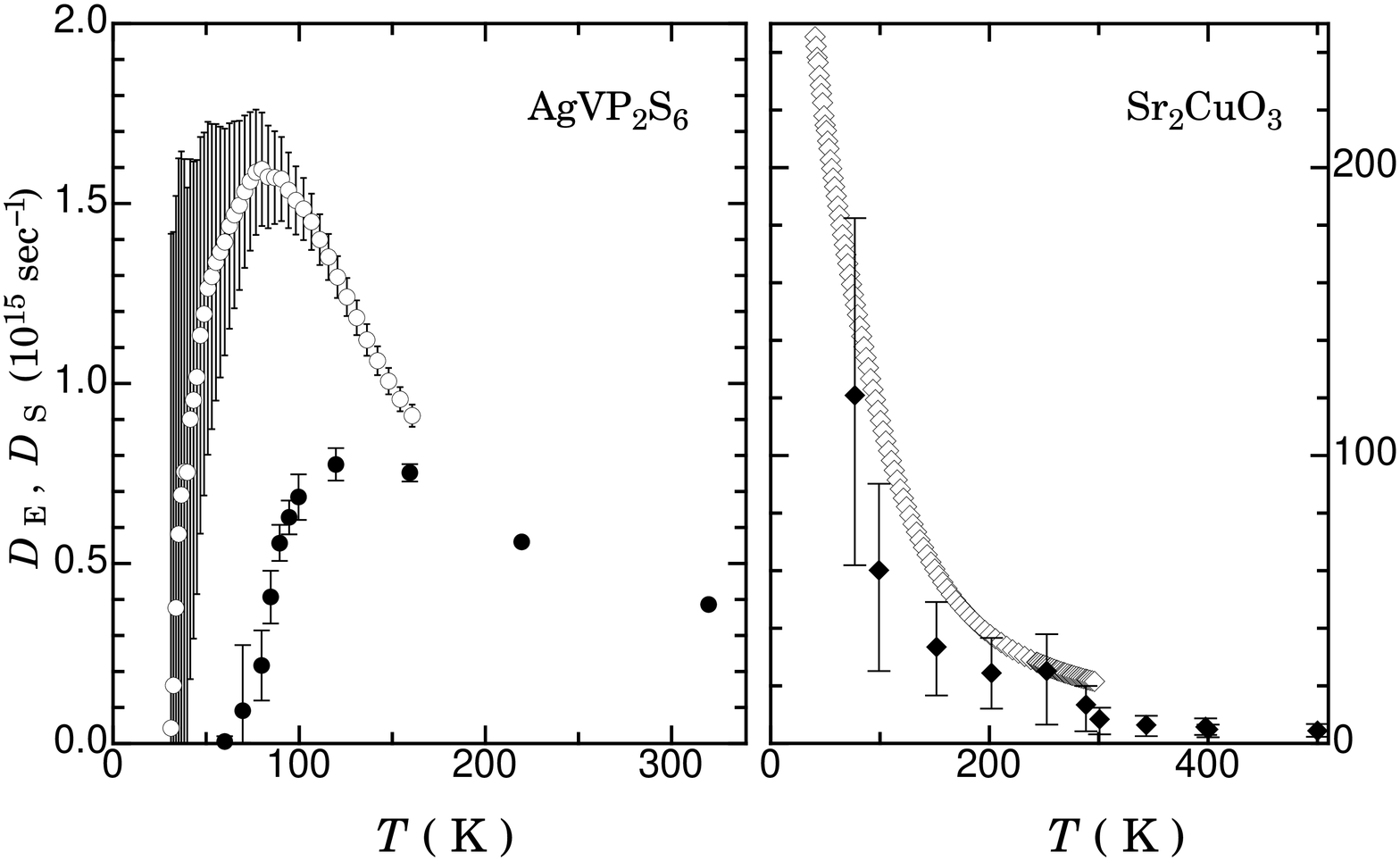,height=2.5in}}
\caption{ The energy diffusion constants $D_E(T)$ (open symbols)
of Sr$_2$CuO$_3$ and AgVP$_2$S$_6$, estimated from thermal
conductivity data,\cite{Sologubenko01,Sologubenko03_AVPS} and the
spin diffusion constants $D_S(T)$ (solid symbols) from NMR
measurements\cite{Thurber01,Takigawa96} (from
Ref.~\onlinecite{Sologubenko03_AVPS}). } \label{DeDs}
\end{figure}

The authors of Ref.~\onlinecite{HeidrichMeisner05} studied the
relation between the thermal Drude weight $D_{\rm th}$ and the
spin Drude weight $K_s$
\begin{equation}\label{eSpinWFL}
\frac{D_{\rm th}}{K_s} = L_0 T,
\end{equation}
which is analogous to the celebrated Wiedemann-Franz law for
electrical conductors. The behavior of the parameter $L_0$ in
different parts of the phase diagram of the  $S=1/2$ XXZ chain
was investigated. The relation between the heat conductivity and
the spin conductivity was tested experimentally in studies of the
$S=1/2$ chain compound Sr$_2$CuO$_3$ and the $S=1$ chain compound
AgVP$_2$S$_6$, respectively. In Fig.~\ref{DeDs}, we show the
comparison of the energy diffusion constant $D_E(T)$ obtained
from the zero-field thermal conductivity
measurements,\cite{Sologubenko01,Sologubenko03_AVPS} and the spin
diffusion constant $D_S(T)$ established by NMR
experiments.\cite{Thurber01,Takigawa96} For Sr$_2$CuO$_3$, both
transport parameters show similar temperature dependences and
differ by a factor close to 2. Also for AgVP$_2$S$_6$, $D_E$ and
$D_S$ adopt similar absolute values and exhibit very similarly
shaped temperature dependences.  Earlier  theoretical
work\cite{Huber69,Huber69a,Lurie74} for $S=1/2$ and $S=1$ chains
and the recent calculations for $S=1$ chains,\cite{Zotos04}
indeed, predict values of the ratio $D_{E}/D_{S}$ in the interval
between  1.4 and 3 at high temperatures.

\section{SUMMARY AND OUTLOOK}

Recent experimental studies accumulated a considerable amount of
evidence for heat transport via magnetic excitations in
paramagnetic one-dimensional spin arrays. In some  cases,
particularly for $S = 1/2$ spin chains, the spin thermal
conductivity is rather large, consistent with theoretical
predictions which claim a ballistic form of transport. Magnetic
defects were identified as one of the important sources for the
scattering of spin excitations. The experiments also suggest that
spin-phonon scattering is very effective in reducing the energy
flow, especially at high temperatures. Recent measurements
provided some support for diffusive heat transport in 1D spin
systems exhibiting a gap in the excitation spectrum.
Nevertheless, the situation is far from being clear, analogous to
a similar controversy in the theoretical sector. Special analyses
of experimental data revealed a link between the heat conduction
and the spin transport in spin-chain systems.

In spite of the experimental progress in certain areas, a number
of problems could not yet be solved. One of the central issues in
current theoretical studies is whether the transport in  spin
chains with a gap in the excitation spectrum is ballistic or
diffusive. The available experimental data sets do not provide
enough solid evidence in favour of either of these expectations.
The number of suitable materials that were studied so far, is
simply too small. For example, only two $S = 1$ chain compounds
were investigated. In both materials, the concentration of
magnetic defects was relatively high, leaving the possibility
that intrinsic features of the thermal conductivity are masked.
Also for the probed spin-ladder materials, the experimentally
observed large heat flow carried by spins does not seem to
contradict the possibility of diffusive transport governed by
intrinsic interactions. In order to clarify this situation, more
experimental work is needed.

Concerning the integrable spin systems with gapless excitation
spectra, the most important problem is to establish, how the spin
thermal conductivity is affected by various possible
integrability-breaking perturbations, such as dimerization,
frustration, weak interchain coupling, next-nearest-neighbor
interactions, anisotropic interactions, various types of
disorder, and coupling to phonons. Very little experimental work
has been done in this direction. Even for the most extensively
studied case of scattering by defects, the $T$ - and
$k$-dependence of the corresponding scattering rate is not yet
established unambiguously. For progress in this direction, model
materials that allow for a controlled alteration of different
perturbations are required.

In spite of the rich variety of phenomena predicted by theory,
very little experimental work has been done on the influence of
magnetic fields on the spin-carried heat conduction in 1D spin
systems. Progress in this area requires materials with weak
intrachain exchange and yet much weaker interchain interactions.
Besides variations in the thermal conductivity, a number of
interesting new features in magnetothermal transport are expected
to be observed.

\section*{ACKNOWLEDGMENTS}
This paper is dedicated to Professor Hilbert von L\"{o}hneysen on the
occasion of his 60th birthday. We acknowledge financial support by
the Deutsche Forschungsgemeinschaft through SFB 608.


\end{document}